\documentstyle[12pt]{article}
\title{\baselineskip=9mm
Role of anharmonicities of nuclear vibrations \\
in fusion reactions at subbarrier energies} 
\author{K.  Hagino$^{1}$, N.  Takigawa$^{1}$, and S.  Kuyucak$^{2}$ \\ \\
\medskip
{\it $^{1}$Department of Physics,
Tohoku University, Sendai 980--77, Japan}
\\
{\it $^{2}$Department of Theoretical Physics, Research School of Physical
Sciences, } \\
{\it Australian National University, Canberra,
ACT 0200, Australia}
}
\date{}

\begin{document}

\maketitle

\begin{center}
{\bf Abstract}
\end{center}

We discuss the effects of double octupole and quadrupole phonon excitations 
in $^{144}$Sm on fusion reactions between $^{16}$O and $^{144}$Sm at 
subbarrier energies.  The effects of anharmonicities of the vibrational 
states are taken into account by using the $sdf$-interacting boson model.  
We compare the results with those in the harmonic limit to show that 
anharmonicities play an essential role in reproducing the experimental 
fusion barrier distribution.  From the analysis of the high quality fusion 
data available for this system, we deduce negative static quadrupole 
moments for both the first 2$^{+}$ and 3$^{-}$ states in $^{144}$Sm.  This 
is the first time that the sign of static quadrupole moments of phonon 
states in a spherical nucleus is determined from the data of subbarrier 
fusion reactions.

\noindent
PACS number(s): 25.70.Jj, 21.60.Fw, 21.10.Ky, 27.60.+j

\newpage

Collective phonon excitations are common phenomena in fermionic many-body 
systems.  In nuclei, low-lying surface oscillations with various 
multipolarities are typical examples.  The harmonic vibrator provides a 
zeroth order description for these surface oscillations, dictating simple 
relations among the level energies and the electromagnetic transitions 
between them.  For example, all the levels in a phonon multiplet are 
degenerate and the energy spacing between neighboring multiplets is a 
constant.  In realistic nuclei, however, there are residual interactions 
which cause deviations from the harmonic limit, e.g., they split levels 
within a multiplet, change the energy spacings, and also modify the ratios 
between various electromagnetic transition strengths.  There are many 
examples of two-phonon triplets ($0^+,2^+,4^+$) of quadrupole surface 
vibrations in even-even nuclei near closed shells.  Though the center of 
mass of their excitation energies are approximately twice the energy of the 
first $2^+$ state, they usually exhibit appreciable splitting within the 
multiplet.  A theoretical analysis of the anharmonicities for the 
quadrupole vibrations was first performed by Brink {\it et al.} 
\cite{BTK65}, where they related the excitation energies of three-phonon 
states to those of double-phonon triplets.  For a long time, however, the 
sparse experimental data on three-phonon states had caused debates on the 
existence of multi-phonon states.  The experimental situation has improved 
rapidly in recent years, and data on multi-phonon states are now available 
for several nuclei.  As a consequence, study of multi-phonon states, and 
especially their anharmonic properties, is attracting much interest 
\cite{CZ96}.

In many even-even nuclei near closed shells, a low-lying 3$^{-}$ excitation 
is observed at a relatively low excitation energy, which competes with the 
quadrupole mode of excitation \cite{BM75}.  These excitations have been 
frequently interpreted as collective octupole vibrations arising from a 
coherent sum of one-particle one-hole excitations between single particle 
orbitals differing by three units of orbital angular momentum.  This 
picture is supported by large E3 transition probabilities from the first 
3$^{-}$ state to the ground state, and suggests the possibility of 
multi-octupole-phonon excitations.  In contrast to the quadrupole 
vibrations, however, so far there is little experimental evidence for 
double-octupole-phonon states.  One reason for this is that E3 transitions 
from two-phonon states to a single-phonon state compete against E1 
transitions.  This makes it difficult to unambiguously identify the 
two-phonon quartet states ($0^+,2^+,4^+,6^+$).  Only in recent years, 
convincing evidences have been reported for double-octupole-phonon states in 
some nuclei, including $^{208}$Pb \cite{YGM96} and $^{144}$Sm \cite{GVB90}.
 
Nuclear surface vibrations have also been studied in connection with 
nuclear reaction problems.  For instance, the influence of nuclear surface 
vibrations on heavy-ion fusion reactions at energies below and near the 
Coulomb barrier has been investigated by many groups (see Ref. \cite{B88} 
for a review).  These studies were later extended to include the effect of 
multi-phonon states \cite{TI86,EL87,KRNR93}.  It has been recognized by now 
as a general phenomenon that such channel couplings cause a significant 
enhancement of fusion cross sections relative to the predictions of 
one-dimensional barrier penetration models \cite{B88}.  Recently, it was 
suggested that the effects of channel couplings can be visualized more 
effectively by studying the second derivative of 
the product of the fusion cross section and the center of mass energy with 
respect to the energy \cite{RSS91}.  This quantity is conventionally called 
{\it fusion barrier distribution}, because it represents the distributed 
fusion barriers induced by the coupling of the relative motion to nuclear 
intrinsic motion in the limit of degenerate spectrum, i.e.  in the limit 
where the excitation energy of the nuclear intrinsic excitation is ignored.  
The excitation function of fusion cross sections has to be measured with 
very high accuracy at small energy intervals in order to deduce meaningful 
barrier distributions from the experimental data.  Thanks to the recent 
developments in experimental techniques \cite{WLH91}, such data are now 
available for several systems, and they have clearly demonstrated the 
sensitivity of the barrier distribution to the details of the channel 
coupling \cite{LDH95}.  For example, the barrier distribution analysis of 
the recently measured accurate data on $^{58}$Ni + $^{60}$Ni fusion 
reaction has shown evidence for coupling of multi-phonon states in 
$^{58}$Ni and $^{60}$Ni \cite{SACN95}.  The barrier distributions were 
shown to be quite sensitive to the number of phonons excited during fusion 
reactions.  This suggests that subbarrier fusion reactions may provide an 
alternative method to identify the existence of multi-phonon states and to 
study their detailed properties such as anharmonicities.

The $^{16}$O + $^{144}$Sm fusion reaction, whose excitation function has 
recently been measured with high accuracy \cite{LDH95}, could serve as a 
test case in this respect.  It has been reported that inclusion of the 
double-phonon excitations of $^{144}$Sm in coupled-channels calculations in 
the harmonic limit destroys the good agreement between the experimental 
fusion barrier distribution and the theoretical predictions obtained when 
only the single-phonon excitations are taken into account \cite{M95}.  On 
the other hand, there are experimental \cite{GVB90,WRZB96} as well as 
theoretical \cite{GS94} support for the existence of the 
double-octupole-phonon states in $^{144}$Sm.  Reconciliation of these 
apparently contradictory facts may be possible if one includes the 
anharmonic effects, which are inherent in most multi-phonon spectra.

The aim of this Letter is to show that the anharmonicities indeed play an 
important role in the fusion reactions between $^{16}$O and $^{144}$Sm.  We 
demonstrate that the anharmonic properties of the quadrupole and octupole 
vibrational excitations in $^{144}$Sm strongly influence the shape of the 
fusion barrier distributions, and lead to a good agreement between the 
experimental data and theoretical predictions.  The excitations of $^{16}$O 
are not included as they are effectively incorporated in the choice of the 
bare potential \cite{HTDHL97b}.  We also estimate the magnitude as well as 
the sign of the quadrupole moments of the quadrupole and octupole 
single-phonon states of $^{144}$Sm from the experimental fusion barrier 
distribution.  

The $sdf$-interacting boson model (IBM) in the vibrational limit provides a 
convenient calculational framework to address these questions \cite{IA87}.  
The vibrational limit of the IBM and the anharmonic vibrator (AHV) in the 
geometrical model are very similar, the only difference coming from 
the finite 
number of bosons in the former \cite{CW88}.  A model for subbarrier fusion 
reactions, which uses the IBM to describe effects of channel couplings, has 
been developed in Ref. \cite{BBK94}.  Following \cite{BBK94}, we assume 
that the Hamiltonian for the fusing system is given by 
\begin{equation}
H=-\frac{\hbar^2}{2\mu}\nabla^2+H_{IBM}+
V_{coup}({\mbox{\boldmath $r$}},\xi),
\end{equation}
where ${\mbox{\boldmath $r$}}$ is the coordinate of the relative motion 
between the projectile and the target, $\mu$ is the reduced mass, and $\xi$ 
represents the internal degrees of freedom of the target nucleus.  $H_{IBM}$ 
is the IBM Hamiltonian for the quadrupole and octupole vibrations in the 
target nucleus, for which we assume the harmonic limit
\begin{equation}
H_{IBM}=\epsilon_d\hat{n}_d + \epsilon_f\hat{n}_f.
\label{hibm}
\end{equation}
Here $\hat{n}_d$ and $\hat{n}_f$ are the number operators for $d$ and $f$ 
bosons, and, $\epsilon_d$ and $\epsilon_f$ are the excitation energies of 
the quadrupole and octupole vibrations, respectively.  Note that we have 
neglected the two-body interactions in Eq. (\ref{hibm}) that give rise to 
anharmonicities in the spectrum.  The reason for this apparently 
self-defeating choice is that anharmonicities in level energies have 
only a marginal effect on the fusion excitation function and the barrier 
distribution.  In fact, our studies show that the fusion barrier 
distribution does not depend so much on the excitation 
energies of the multi-phonon states once the energies of the single-phonon 
quadrupole and octupole states are fixed.  As we will see later, the main 
effects of the anharmonicity on fusion barrier distributions come from the 
deviation of the transition probabilities from the harmonic limit.

The coupling between the relative motion and the intrinsic motion of the 
target nucleus is described by $V_{coup}$ in Eq. (1), which consists of the 
Coulomb and nuclear parts.  Following Ref. \cite{BBK94}, and using 
the no-Coriolis approximation \cite{TI86}, they are given by
\begin{eqnarray}
&&V_C(r,\xi)=\frac{Z_PZ_Te^2}{r}\left(1
+ \frac{3}{5}\frac{R_T^2}{r^2} \frac{\beta_2 \hat{Q}_{20}}{\sqrt{4\pi N}}
+ \frac{3}{7}\frac{R_T^3}{r^3} \frac{\beta_3 \hat{Q}_{30}}{\sqrt{4\pi N}} 
\right) \label{vc}, \nonumber \\
&&V_N(r,\xi)=-V_0 \left[ 1+\exp \left(\frac{1}{a} \left( r-R_0 - R_T ( 
\beta_2 \hat{Q}_{20} + \beta_3 \hat{Q}_{30})/ \sqrt{4\pi N} \right) 
\right)\right]^{-1}.
\label{v}
\end{eqnarray}
Here, $N$ is the boson number, the subscripts $P$ and $T$ refer to the 
projectile and target nuclei, respectively, and $R_0=R_P + R_T$.  The 
scaling of the coupling strength with $\sqrt{N}$ is introduced to ensure 
the equivalence of the IBM and the geometric model results in the large $N$ 
limit \cite{BBK94}.  Further, $\beta_2$ and $\beta_3$ in Eq. (\ref{v}) are the 
quadrupole and octupole deformation parameters, which are usually estimated 
from the electric transition probabilities using the expression 
$\beta_{\lambda}=4\pi(B(E\lambda)\uparrow)^{1/2}/3Z_T e R_T^\lambda$.  
However, this formula does not hold for anharmonic vibrators. Therefore, we 
treat $\beta_2$ and $\beta_3$ as free parameters and look for their optimal 
values to reproduce the experimental data.  Finally, $\hat{Q}_2$ and 
$\hat{Q}_3$ in Eq. (\ref{v}) are the quadrupole and the octupole operators in 
the IBM, which we take as
\begin{eqnarray}
&&\hat{Q}_2=s^{\dagger}\tilde{d} + sd^{\dagger} +
\chi_2(d^{\dagger}\tilde{d})^{(2)}
+ \chi_{2f}(f^{\dagger}\tilde{f})^{(2)}, \nonumber \\ 
&&\hat{Q}_3=sf^{\dagger} + 
\chi_3(\tilde{d}f^{\dagger})^{(3)} + h.c., 
\label{op}
\end{eqnarray}
where tilde is defined as $\tilde{b}_{l\mu}=(-)^{l+\mu}b_{l-\mu}$.  When 
all the $\chi$ parameters in Eq.~(\ref{op}) are zero, quadrupole moments of 
all states vanish, and one obtains the harmonic limit 
in the large $N$ limit. 
Non-zero values of $\chi$ generate quadrupole moments and are responsible 
for the anharmonicities in electric transitions.

Our coupled-channel calculations include a number of new features that 
improve on previous calculations.  
We do not employ the ``constant coupling'' approximation, 
which is often introduced in simplified calculations. 
Another important aspect of our formalism is that we do not 
introduce the usual linear coupling approximation by expanding the nuclear 
part in Eq. (\ref{v}) with respect to the deformation parameters, but we keep 
the couplings to the intrinsic motion to all orders.  The full order 
treatment is crucial in order to quantitatively, as well as qualitatively, 
describe heavy-ion subbarrier fusion reactions \cite{BBK94,HTDHL97a}.  Also, 
we take into account the finite excitation energies in the target nucleus, 
which have been neglected in previous applications of the IBM to subbarrier 
fusion reactions \cite{BBK94}.  Clearly, excitation energies of the order 
of 1 MeV, as typically encountered in vibrational nuclei, are too large to be 
ignored in fusion dynamics.

The model parameters are determined as follows.  The standard prescription 
for boson number (i.e.  counting pairs of nucleons above or below the 
nearest shell closure) would give $N=6$.  However, it is well known that 
the effective boson numbers are much smaller due to the $Z=64$ subshell 
closure \cite{CW88}.  The suggested effective numbers in the literature 
vary between $N=1$ and 3.  We adopted $N=2$ in our calculations, since 
there are experimental signatures for the two-phonon states, but no 
evidence for three-phonon states in $^{144}$Sm.  The parameters of the IBM 
Hamiltonian Eq.~(\ref{hibm}) are simply determined from the excitation 
energies of the first $2^+$ and $3^-$ states in $^{144}$Sm as 
$\epsilon_d=1.66$ MeV and $\epsilon_f=1.81$ MeV. The nuclear potential 
parameters are taken from the exhaustive study of this reaction in 
Ref.~\cite{M95} as $V_0 = 105.1$ MeV, $R_0 = 8.54$ fm and $a = 0.75$ fm.  
Finally, the target radius is taken to be $R_T = 5.56$ fm.

The results of the coupled-channels calculations are compared with the 
experimental data in Fig.~1.  The upper and the lower panels in Fig.~1 show 
the excitation function of the fusion cross section and the fusion barrier 
distributions, respectively.  The experimental data are taken from 
Ref.~\cite{LDH95}.  The dotted line is the result in the harmonic limit, 
where couplings to the quadrupole and octupole vibrations in $^{144}$Sm are 
truncated at the single-phonon levels.  The deformation parameters are 
estimated to be $\beta_2$=0.11 and $\beta_3$=0.21 from the electric 
transition probabilities.  The dotted line reproduces the experimental data 
of both the fusion cross section and the fusion barrier distribution 
reasonably well, though the peak position of the fusion barrier 
distribution around $E_{cm}=65$ MeV is slightly shifted.  As was shown in 
Ref.~\cite{M95}, the shape of the fusion barrier distribution becomes 
inconsistent with the experimental data when the double-phonon channels are 
included in the harmonic limit (the dashed line).  The good agreement is 
recovered when one takes the effects of anharmonicity of the vibrational 
motion into account.  These results are shown in Fig.~1 by the solid line.  
This calculation has been performed using the parameters, $\beta_2=0.13$, 
$\beta_3=0.23$, $\chi_2=-3.30$, $\chi_{2f}=-$2.48, and $\chi_3$=2.87, which 
are obtained from a $\chi^2$ fit to the fusion cross sections.  The 
$\chi^2$ fit gave a unique result, regardless of the starting values.  The 
non-zero $\chi$ values indicate the anharmonic effects in the transition 
operators.  The slight change in the values of the deformation parameters 
from those in the harmonic limit results from the renormalization effects 
due to the extra terms in the operators given in Eq.~(4).  Note that the 
solid line agrees with the experimental data much better than the dotted 
line.

One of the pronounced features of an anharmonic vibrator is that the 
excited states have non-zero quadrupole moments \cite{BM75}.  Using the 
$\chi$ parameters extracted from the analysis of fusion data in the E2 
operator, $T$(E2$)=e_B \hat Q_2$, we can estimate the static quadrupole 
moments of various states in $^{144}$Sm.  Here, $e_B$ is the effective 
charge, which is determined from the experimental $B($E2$;0\to 2^+_1)$ 
value as $e_B = 0.16~eb$.  For the quadrupole moment of the first 2$^+$ 
and 3$^-$ states, 
we obtain $-$0.28 b and $-$0.70 b, respectively.
The negative sign of the quadrupole moment of the octupole-phonon state is 
consistent with that suggested from the neutron pick-up reactions on 
$^{145}$Sm \cite{KGE89}.

In the case of rotational coupling, fusion barrier distributions strongly 
depend on the sign of the quadrupole deformation parameter through the 
reorientation term.  Also, it has been reported that fusion barrier 
distributions are very sensitive to the sign of the hexadecapole 
deformation parameter \cite{LLW93}.  Similarly, it is likely that the shape 
of fusion barrier distributions changes significantly when one inverts the 
sign of the quadrupole moment in a spherical target.  Fig.~2 shows the 
influence of the sign of the quadrupole moment of the excited states on the 
fusion cross section and the fusion barrier distribution.  The solid line 
is the same as in Fig.~1 and corresponds to the optimal choice for the 
signs of the quadrupole moments of the first 2$^+$ and 3$^-$ states.  The 
dotted and dashed lines are obtained by changing the sign of the $\chi_2$ 
and $\chi_{2f}$ parameters in Eq.~(\ref{op}), respectively, while the 
dot-dashed line is the result where the sign of both $\chi_2$ and 
$\chi_{2f}$ parameters are inverted.  The change of sign of $\chi_2$ and 
$\chi_{2f}$ is equivalent to taking the opposite sign for the quadrupole 
moment of the excited states.  Fig.~2 demonstrates that subbarrier fusion 
reactions are indeed sensitive to the sign of the quadrupole moment of 
excited states.  The experimental data are reproduced only when the correct 
sign of the quadrupole moment are used in the coupled-channels 
calculations.  Notice that the fusion excitation function is completely 
insensitive to the sign of the quadrupole moment of the first $2^+$ state, 
but strongly depends on that of the first $3^-$ state.  In contrast, the 
fusion barrier distribution can probe the signs of the quadrupole moments 
of both the first $2^+$ and $3^-$ states.  This study shows that the sign of 
quadrupole moments in spherical nuclei can be determined from subbarrier 
fusion reactions, especially through the barrier distribution.

In summary, we have analyzed the experimental fusion excitation function 
for $^{16}$O + $^{144}$Sm reaction with a model which explicitly takes into 
account the effects of anharmonicity of the vibrational modes of excitation 
in $^{144}$Sm.  We have focused on the anharmonic effects of the 
phonon excitations in $^{144}$Sm and found that the best fit to the 
experimental data requires negative quadrupole moments for the first 2$^+$ 
and the first 3$^-$ states.  As a general conclusion, we find that 
heavy-ion subbarrier fusion reactions, and in particular, barrier 
distributions extracted from the fusion data, are very sensitive to the 
sign of the quadrupole moments of phonon states in the target nucleus.

\bigskip

The authors thank J.R. Leigh, M. Dasgupta, D.J. Hinde, and J.R. Bennett for 
useful discussions.  K.H. and N.T. also thank the Australian National 
University for its hospitality and for partial support for this project.  
The work of K.H. was supported by the Japan Society for the Promotion of 
Science for Young Scientists.  This work was supported by the Grant-in-Aid 
for General Scientific Research, Contract No.06640368 and No.08640380, and 
the Grant-in-Aid for Scientific Research on Priority Areas, Contract 
No.05243102 and 08240204 from the Japanese Ministry of Education, Science 
and Culture, and a bilateral program of JSPS between Japan and Australia.

\newpage

\begin{center}
{\bf Figure Captions}
\end{center}

\noindent {\bf Fig.~1:} Comparison of the experimental fusion cross section 
(the upper panel) and fusion barrier distribution (the lower panel) with 
the coupled-channels calculations for $^{16}$O + $^{144}$Sm reaction.  The 
experimental data are taken from Ref.~\cite{LDH95}.  The solid line shows 
the results of the present IBM model including the double-phonon states and 
anharmonic effects.  The dotted and the dashed lines are the results of the 
single- and the double-phonon couplings in the harmonic limit, 
respectively.

\noindent
{\bf Fig.~2:}
Dependence of the fusion cross section and barrier distribution on the 
sign of the quadrupole moment of the excited states in $^{144}$Sm.  The 
meaning of each line is indicated in the inset.

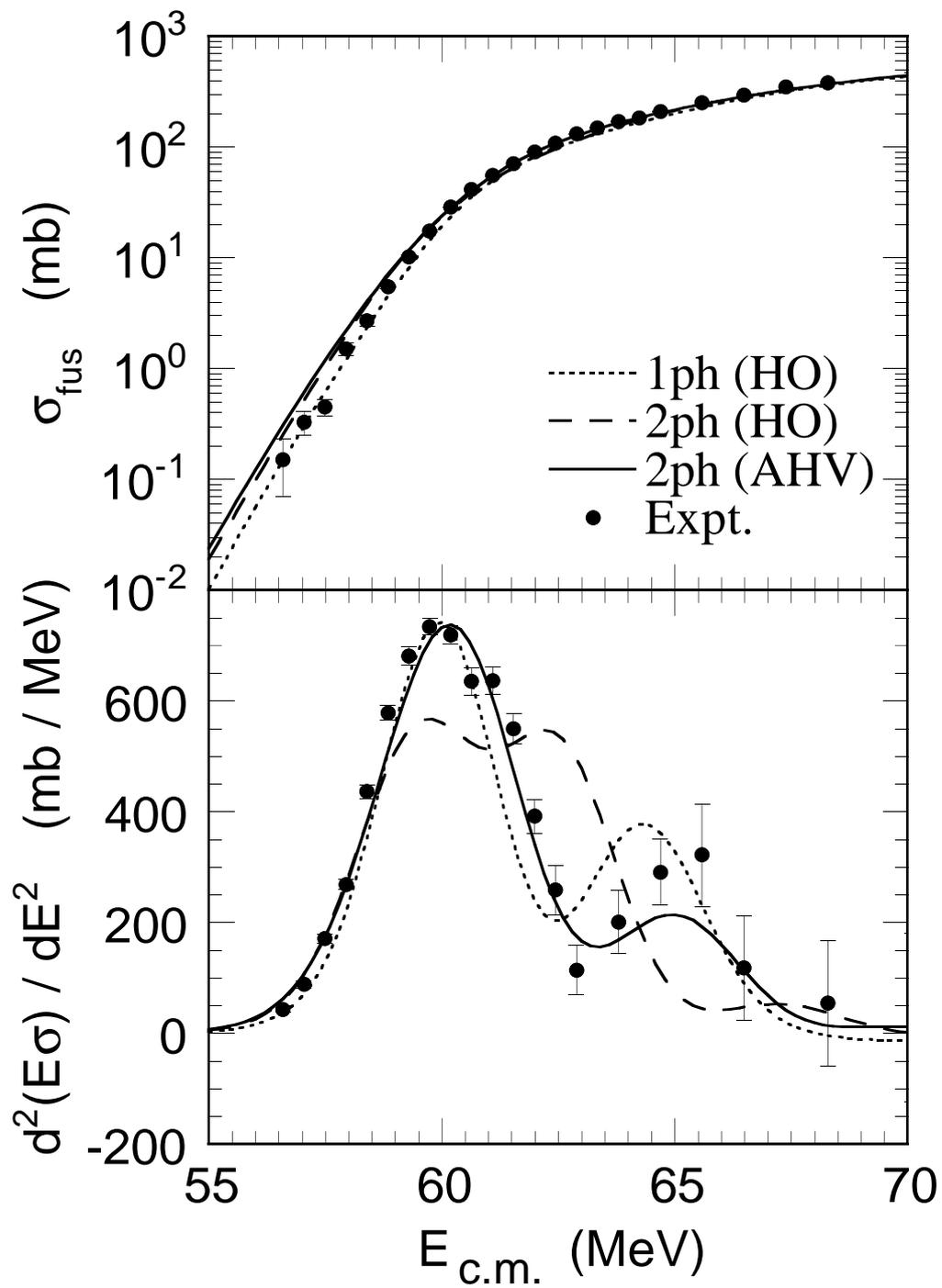
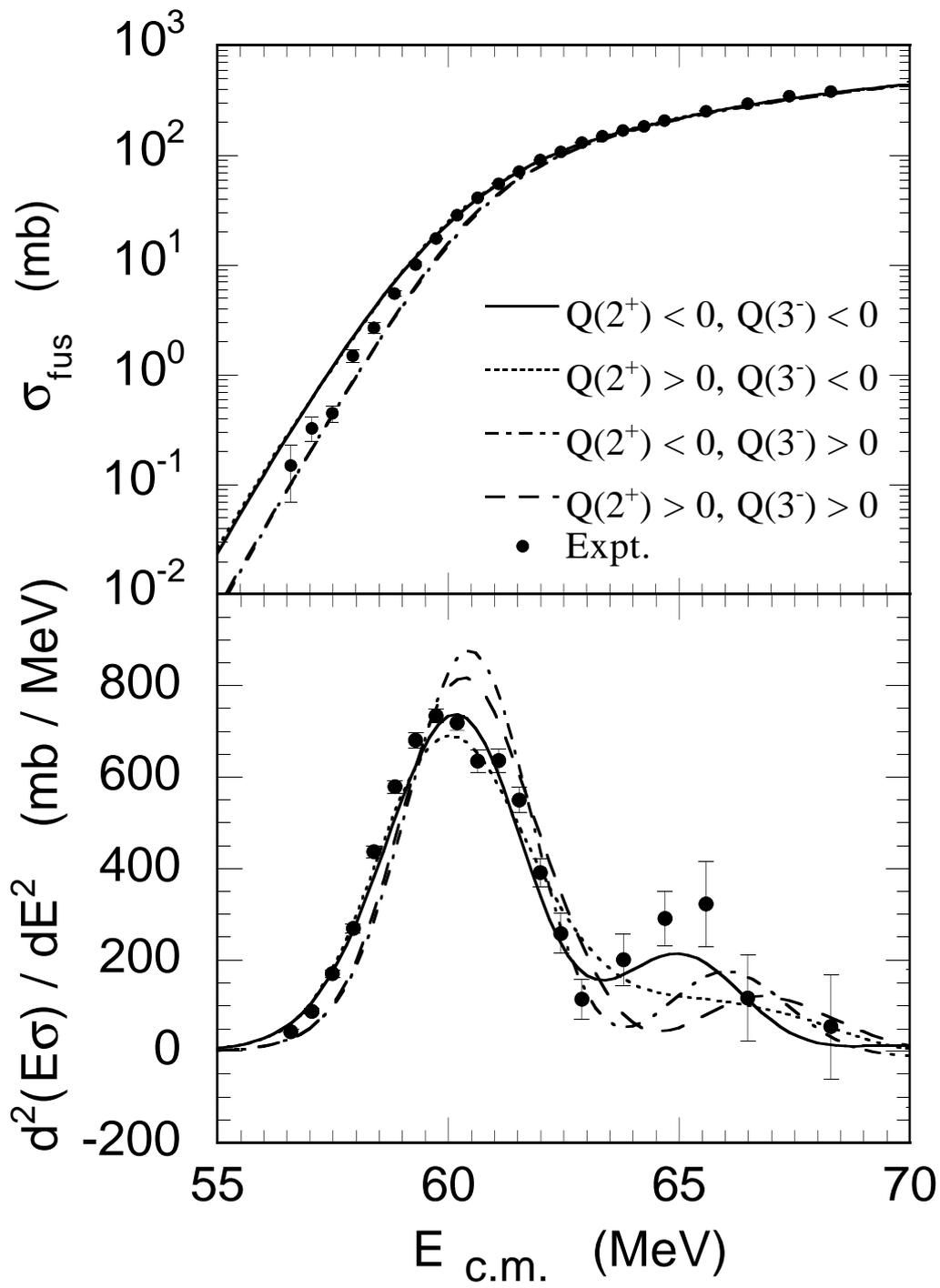
\end{document}